# On the contribution of nearly-critical spin and charge collective modes to the Raman spectra of high-$T_c$ cuprates

S. Caprara[a]*, C. Di Castro[a], T. Enss[b], and M. Grilli[a]

[a]*Dipartimento di Fisica, Università di Roma "La Sapienza", Piazzale Aldo Moro, 2 – 00185 Roma - Italy*
[b]*Physik-Department, Technische Universität München, James-Franck-Straße, 85747 Garching - Germany*

**Abstract**

We discuss how Raman spectra are affected by nearly-critical spin and charge collective modes, which are coupled to charge carriers near a stripe quantum critical point. We show that specific fingerprints of nearly-critical collective modes can indeed be observed in Raman spectra and that the selectivity of Raman spectroscopy in momentum space may also be exploited to distinguish the spin and charge contribution. We apply our results to discuss the spectra of high-$T_c$ superconducting cuprates finding that the collective modes should have masses with substantial temperature dependence in agreement with their nearly critical character. Moreover spin modes should be more diffusive than charge modes indicating that in stripes the charge is nearly ordered, while spin modes are strongly overdamped and fluctuate with high frequency.

## 1. Introduction

Many experiments [1] and theoretical predictions [2] indicate that the physics of high-$T_c$ cuprates might be controlled by the proximity to a quantum critical point. One of the candidates for the corresponding ordered phase is a finite-wavelength charge and spin density modulation, possibly in the form of stripe-like textures [3]. The question then arises whether and how fingerprints of nearly-critical collective modes (CMs) are seen in various experiments, thus revealing the incipient critical behavior.

Recently, the lack of an evident scaling in the far-infrared optical spectra of the cuprates was interpreted as evidence against standard quantum criticality [4]. However, optical conductivity might be severely conditioned by momentum conservation, and should indeed vanish in a clean Galilean-invariant system, unless the fermion charge carriers are coupled to (dynamical) degrees of freedom which dissipate the current. If these degrees of freedom are

―――――――――――
* Corresponding author. Tel.: +039-0649914294; fax: +39-064957697.
  E-mail address: sergio.caprara@roma1.infn.it.

nearly critical CMs, the momentum dissipation mechanism introduces non-universal frequency scales, leaving the way open to a variety of phenomena, ranging from partial scaling of the optical spectra, to the absence of any evident feature to be associated with quantum criticality, in qualitative agreement with the variety of experimental findings [5,6]. On the contrary, Raman spectroscopy is not constrained by momentum conservation. Moreover, the Raman vertices of various symmetries have a characteristic momentum dependence, which might directly probe different regions of the quasiparticle Fermi surface, giving precise indications on the typical wavevectors of the CMs. This characteristic of Raman spectroscopy allows us to extract both from the anomalous dispersing peaks and from the underlying background which are experimentally observed in the underdoped region of the phase diagram of the cuprates, precise information about the nature of the relevant CMs. Preliminary results seem to indicate that our model may provide a simple interpretation for the specific temperature, doping, and symmetry dependence of Raman spectra in cuprates [5,7].

## 2. Raman spectra of fermions coupled to CMs

We calculate Raman spectra within an effective model of fermion quasiparticles (the charge carriers) coupled to nearly-critical CMs, which is apt to describe the physics of high-$T_c$ cuprates near a stripe instability [8], considering the diagrams of Fig. 1.

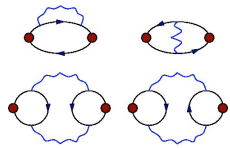

Fig. 1. Set of diagrams adopted in this work to calculate the Raman response function. The symbols are explained in the text.

The wavy blue line represents a CM Matsubara propagator

$$D(\mathbf{q},\omega_\ell) = \left[ m + \nu \left| \mathbf{q} - \mathbf{q}_c \right|^2 + \left|\omega_\ell\right| + \frac{\omega_\ell^2}{\overline{\omega}} \right]^{-1},$$

obtained expanding a RPA-dressed interaction among the fermionic quasiparticles, with a long- and short-range static repulsion, and a frequency dependent phonon-mediated term [9], at low Matsubara frequency $\omega_\ell$ and for wavevectors $\mathbf{q}$ near the critical wavevector $\mathbf{q}_c$ [5-8]. The CM mass $m$ measures the distance form criticality and the parameter $\nu$ measures the CM dispersion. Apart from the $\omega_\ell^2$ term, the propagator describes a diffusive CM, with a finite characteristic wavevector $\mathbf{q}_c$. We consider here two CMs which should be relevant near the stripe instability in the cuprates, charge fluctuations, with typical wavevectors $\mathbf{q}_c \approx (\pm\pi/2, 0)$ and $(0, \pm\pi/2)$, and antiferromagnetic spin fluctuations, with typical wavevectors $\mathbf{q}_c \approx (\pi,\pi)$. The $\omega_\ell^2$ term arises from the frequency dependence of the phonon-mediated interaction and the related frequency scale $\overline{\omega}$ entails fermion momentum dissipation, if other mechanisms breaking the Galilean invariance, such as impurities or Umklapp, are absent [6]. This dissipation channel is suppressed for $\overline{\omega} \to \infty$. For frequencies larger than $\overline{\omega}$ the CM changes from diffusive to propagating.

The red dots in Fig. 1 represent Raman vertices, which depend on the fermion band curvature and on the polarization of the ingoing and outgoing photon. The solid black lines with arrows indicate the fermion Matsubara propagator

$$G(\mathbf{k},\varepsilon_n) = [i\varepsilon_n - \xi(\mathbf{k})]^{-1},$$

where $\xi(\mathbf{k})$ is the fermion band dispersion.

The first two diagrams in Fig. 1 are the usual self-energy and vertex (S-V) contributions, whereas the other two diagrams are similar to the diagrams considered by Aslamazov and Larkin (AL) in the theory of paraconductivity in superconductors. For the conductivity they should not be neglected in order to preserve the gauge invariance of the calculation [6,10]. In Raman spectra the two sets, S-V and AL, correspond to physically different processes and require different resummation schemes.

– *The S-V contribution* –

Owing to their different physical character in Raman, we separate in the following discussion the two contributions, starting with the S-V term. We accordingly define a non-resummed Raman response $\tilde{\chi}_{S-V}$ from the first two diagrams of Fig. 1. The presence of impurities can be described introducing the memory function [5] $M(\omega) = i\Gamma - \omega W^{-1}\tilde{\chi}_{S-V}$, where $W$ is the optical weight and $\Gamma$ is the impurity scattering rate. The full Raman response function is then found as

$$\chi_{S-V} = \frac{W\omega}{\omega + M_{S-V}(\omega)}. \quad (1)$$

The symmetry of the momentum-dependent Raman vertex, which can be filtered by detecting polarized ingoing and outgoing phonons, selects the CMs according to their characteristic wavevector. In the following we focus on $B_{1g}$ and $B_{2g}$ symmetry, to connect with the experiments on cuprates. The S-V contribution in $B_{1g}$ symmetry, with a vertex $\gamma = \cos k_x - \cos k_y$, vanishes for charge CMs with typical wavevectors $\mathbf{q}_c \approx (\pm\pi/2, 0)$ and $\approx (0, \pm\pi/2)$, and is finite for antiferromagnetic spin fluctuations with $\mathbf{q}_c \approx (\pi, \pi)$. The opposite is true in the $B_{2g}$ symmetry, with a vertex $\gamma = \sin k_x \sin k_y$. Therefore, as far as the S-V contribution is concerned, the two CMs are probed *separately* of each other, in two different channels.

For each symmetry, the S-V Raman response function of the corresponding CM can be written as in Eq. (1). The memory function can be cast in the Allen-approximation form [11]

$$\mathrm{Im}\, M_{S-V}(\omega) = \frac{g^2}{\omega}\int_0^\infty dz\, \alpha^2 F(z) \left[2\omega \coth\left(\frac{z}{2T}\right) - (z+\omega)\coth\left(\frac{z+\omega}{2T}\right) + (z-\omega)\coth\left(\frac{z-\omega}{2T}\right)\right],$$

where $T$ is the temperature, $g$ is a symmetry-dependent coupling between the CM and the fermions, and the CM spectral strength is

$$\alpha^2 F(\omega) = \arctan\left(\frac{\Lambda\overline{\omega} - \omega^2}{\overline{\omega}\omega}\right) - \arctan\left(\frac{m\overline{\omega} - \omega^2}{\overline{\omega}\omega}\right),$$

where the ultraviolet cutoff $\Lambda$ of the collective-mode band dispersion has been introduced. The real part of $M_{S-V}$ is found by Kramers-Kronig transformation. Our result generalizes the so-called gapped marginal Fermi liquid previously considered in the context of optical spectra [12], to which it reduces in the limiting case of small $\overline{\omega}$, with fixed $\Lambda\overline{\omega}$ and $m\overline{\omega}$.

Within our model, the two scales which define the crossover regime from low- to high-frequency can be read off $\alpha^2 F(\omega)$, and are $\omega_1 = m\overline{\omega}$, and $\omega_2 = \Lambda\overline{\omega}$. The first is determined by the CM mass, and the second by the cutoff $\Lambda$, and both depend on the scale $\overline{\omega}$, marking the crossover from a diffusive to a propagating CM. The inset in Fig. 2 reports three typical spectral densities for three different values of the parameter $\alpha \equiv 1/\overline{\omega}$. Apparently, the smaller is $\alpha$, and the more diffusive is the CM, with a larger weight at low frequency.

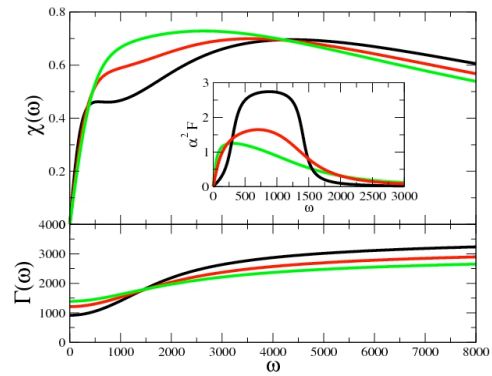

Fig. 2: Typical Raman spectra (Upper panel) and scattering rate (lower panel) due to S-V diagrams with the CM having $\omega_2 = 1400$ cm$^{-1}$, $\omega_1 = 100$ cm$^{-1}$, $\overline{\omega} = 200$ cm$^{-1}$, (black curve), $\overline{\omega} = 2000$ cm$^{-1}$ (green curve), and $\omega_2 = 1400$ cm$^{-1}$, $\omega_1 = 330$ cm$^{-1}$, $\overline{\omega} = 1000$ cm$^{-1}$ (red curve). Inset: CM spectral strength. The parameters are the same as in the main panels.

In Fig. 2, we report different shapes of the Raman spectra (upper panel) and the related $\Gamma(\omega)$ (lower panel). It is evident that a more variable $\Gamma(\omega)$ produces a step-hump spectrum, (black curves) while a more "flattish" scattering rate results in smoother spectra (red and green curves).

The shape of the calculated Raman response function is essentially determined by $\mathrm{Im}\,M_{S-V}$, which acts as an inverse lifetime $\Gamma(\omega)$. If $\Gamma(\omega)$ does not vary in a sizable way, a Drude-like Raman response is observed, with a peak of frequency and width of order $\Gamma(\omega)$ (cf. green curve of Fig. 2). Instead, if $\Gamma(\omega)$ increases substantially, the calculated Raman spectra exhibits two structures, a smaller Drude-like peak corresponding to the low-frequency value of $\Gamma(\omega)$, and a broad bump which can be interpreted as a broader Drude-like peak roughly at the high-frequency value of $\Gamma(\omega)$. The two structures appear clearly distinct if the low- and high-frequency values of $\Gamma(\omega)$ are sizably different (black curves in Fig. 2), otherwise the low-frequency structure may appear as a shoulder of the broader structure (red curves in Fig. 2). In turn, $\Gamma(\omega)$ is determined by the CMs scattering the fermions. Specifically a more propagating CM, with $\omega_1 = m\bar{\omega}$ larger than $T$, gives small scattering at low frequencies and $\Gamma(\omega)$ starts small. At larger frequencies the CM spectrum flattens, $\Gamma(\omega)$ becomes linear and it produces a flat (marginal-FL-like) Raman response. Once $\Gamma(\omega)$ saturates to a large value at high frequency, the Raman response becomes that of a Drude-like term with a large scattering rate. Therefore, if the CM is essentially propagating, $\bar{\omega}$ is small, and the spectral strength of the CM clearly separates the two scales, $\Gamma(\omega)$ varies in a significant way, and the two features discussed above appear as clearly distinct. If, on the other hand, the CM is more diffusive, $\bar{\omega}$ is larger, $\alpha^2 F(\omega)$ is smoother, and does not clearly separate the two scales. In this case, $\Gamma(\omega)$ does not vary in a significant way with frequency, and the two features discussed above are not clearly distinguishable.

– *The AL contribution* –

The AL-like diagrams in Fig. 1, instead, give a finite contribution only in the case of the charge CM and only in the $B_{1g}$ symmetry, and vanish for charge CMs in the $B_{2g}$ symmetry and for antiferromagnetic spin fluctuations in all symmetries[7]. The AL-like contribution due to charge CMs adds to the contribution of antiferromagnetic spin CMs, and appears as an additional low-frequency peak in the $B_{1g}$ symmetry, which is weak if the collective-mode mass $m$ is large, and becomes softer and more pronounced upon reducing $m$, i.e., when approaching criticality [5,7].

## 3. Raman spectra of the cuprates

Already at a first glance, the Raman spectra of cuprates display several peculiarities, which depend on the symmetry of the Raman vertices [13,14]. We here comment on the results of a systematic analysis of LSCO materials at different dopings ranging from 0.02 to 0.26 [15].

The $B_{2g}$ spectra, in the frequency range 0÷8000 cm$^{-1}$, are characterized by a step-like shape at low frequency ($\omega \leq 2000$ cm$^{-1}$) and then display a hump at $\omega \approx 5000 \div 6000$ cm$^{-1}$, at least twice as high as the lower-frequency step. The $B_{1g}$ spectra, in the same frequency range, have instead a more rounded shape at low $\omega$, whose height is comparable with the height of the hump, so that the step may appear as a shoulder of the high-frequency hump.

Our theoretical analysis first of all shows that the two spectra must be related to two distinct CMs with different and finite characteristic wavevector $\mathbf{q}_c$ (otherwise the $B_{1g}$ and $B_{2g}$ spectra would look much more similar). Natural candidates are the spin and charge fluctuations near a stripe instability. As we have shown, the former, give a S-V contribution to $B_{1g}$ spectra, while the latter, contribute to the $B_{2g}$ spectra. Furthermore, the previous analysis also allows to infer that spin CMs are more diffusive to produce the more rounded spectra of the $B_{1g}$ channel. On the other hand, the step+hump shape of the $B_{2g}$ spectra indicates that, in optimally doped and underdoped LSCO, charge CMs should be much less diffusive (i.e., with $\bar{\omega} \approx 100 \div 200$ cm$^{-1}$). This might indicate that charge is more ordered in these systems and its fluctuations are less damped and slower than spin fluctuations.

As far as the $T$ dependence is concerned, a common feature of the spectra calculated with fixed parameters is a substantial temperature dependence (both at low and high frequencies) in agreement with the bosonic character of the interaction mediators. This substantial dependence is instead absent in the experimental Raman spectra and our analysis successfully accounts for this apparent discrepancy. We first concentrate on the low frequency part of the spectrum ($\omega \leq 1000$ cm$^{-1}$). In this frequency region we found in a previous work [7] evidence for a critical behavior of CMs from the anomalous peak, which appears at low frequency in the $B_{1g}$ symmetry [14]. This peak can be explained by an AL-like contribution due to charge CMs. Once the charge CM mass $m$ is adjusted to fit the experimental spectra, we find that $m$ is reduced with decreasing $T$, as if criticality was approached, and then crosses over to a regime of small finite mass down to the lowest $T$, indicating that charge order might be prevented by other mechanisms, such as disorder and/or incipient pairing leading to superconductivity. We therefore borrowed the temperature dependence of the charge CM mass to investigate the $T$ dependence of the $B_{2g}$ spectra, where these CMs should contribute. In this way, we remarkably find that the spectra lose much of their *T* dependence in the whole frequency range, because the increase of scattering with increasing temperature is compensated by the larger CM mass. Fig. 3 reports for illustrative purpose a comparison between spectra calculated at a fixed CM mass $m$=200 cm$^{-1}$ and spectra obtained with a temperature dependence of the mass, similar to the one obtained to fit the anomalous AL absorption in the $B_{1g}$ channel. Clearly the spectra obtained with a *T*-dependent mass (right panel) look much more temperature independent (as in experiments).

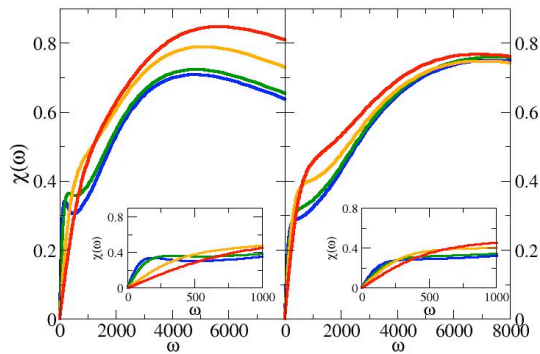

Fig. 3: Typical $B_{2g}$ Raman spectra due to S-V diagrams at different temperatures [T=50 (blue), T=90 (green), T=200 (orange) and T=300 (red) (in cm$^{-1}$)] with fixed charge CM mass $m$=200 cm$^{-1}$, (left panel) and with T-dependent mass [$m(50)=10$ cm$^{-1}$; $m(90)=30$ cm$^{-1}$; $m(200)=350$ cm$^{-1}$; $m(300)=600$ cm$^{-1}$]. Insets: Same spectra on a shorter frequency scale

## 4. Concluding remarks

Our preliminary analysis shows that Raman spectra can provide valuable information about the nature and characteristics of the relevant CMs, which dominate the physics of the cuprates. Our theory seems then able to account for the main features of the experimental spectra, as well as for the peculiar temperature and doping behavior of the low-frequency spectra. The main ingredients are strongly diffusive spin CMs and more propagating charge CMs. The spin CMs have rather large typical frequencies, in agreement with inelastic neutron scattering experiments in LSCO [16], while the charge CMs have smaller typical frequencies. To account for the *T* dependence of the spectra, a nearly-critical character of the CMs has to be assumed, with substantial temperature dependence of their mass.

A thorough survey of the existing experimental data on the Raman spectra of high-T$_c$ cuprates is now in progress [17].

## Acknowledgements

We acknowledge stimulating discussions with C. Castellani, J. Lorenzana, R. Hackl, T. Devereaux. We acknowledge financial support from MIUR-PRIN 2005 prot. 2005022492. SC acknowledges the kind hospitality of the Donostia International Physics Center, where part of this work was carried out.